\documentclass{article}
\usepackage{spconfa4,amsmath,graphicx}
\usepackage{multirow}
\usepackage{graphicx}
\usepackage{makecell}
\usepackage{subfig}
\usepackage{float}
\usepackage{dutchcal}
\usepackage{amssymb}
\title{monaural speech enhancement on drone via Adapter based transfer learning}

\name{Xingyu Chen, Hanwen Bi, Wei-Ting Lai, Fei Ma
}
\address{Audio and Acoustic Signal Processing Group, Australian National University}

\begin{document}
%
\maketitle
\begin{abstract}
Monaural Speech enhancement on drones is challenging because the ego-noise from the rotating motors and propellers leads to extremely low signal-to-noise ratios at onboard microphones. 
Although recent masking-based deep neural network methods excel in monaural speech enhancement, they struggle in the challenging drone noise scenario. 
Furthermore, existing drone noise datasets are limited, causing models to overfit. 
Considering the harmonic nature of drone noise, this paper proposes a frequency domain bottleneck adapter to enable transfer learning. 
Specifically, the adapter's parameters are trained on drone noise while retaining the parameters of the pre-trained Frequency Recurrent Convolutional Recurrent Network (FRCRN) fixed.
Evaluation results demonstrate the proposed method can effectively enhance speech quality. 
Moreover, it is a more efficient alternative to fine-tuning models for various drone types, which typically requires substantial computational resources.

\end{abstract}
\begin{keywords}
drone audition, speech enhancement, single-channel, ego-noise reduction, fine-tuning
\end{keywords}
\section{Introduction}
\label{sec:intro}

Speech enhancement using a drone-mounted microphone enables services in diverse areas, such as search and rescue missions, video capture, and filmmaking~\cite{basiri2012robust}.
However, the microphone's proximity to the drone's noise sources, such as motors and propellers, results in an extremely low signal-to-noise ratio (SNR) of recordings, typically ranging from -25 to -5 dB~\cite{mukhutdinov2023deep}. This severely limits the drone audition applications.

While multi-channel microphone array methods are commonly used to improve audio quality, 
they often require specialized hardware that is either too large or heavy for most drones.
For instance, drones used in the DREGON dataset~\cite{strauss2018dregon} are equipped with 
an 8-channel microphone array and have a total weight of up to 1.68 kg, and Wang \textit{et al.}~\cite{wang2016ear} employ a drone with a 0.2 m diameter circular microphone array.
Moreover, the performance of these methods degrades significantly in dynamic scenarios with moving microphones or sound sources~\cite{wang2018acoustic}. 
Thus, developing efficient single-channel solutions is preferred for extending the 
applicability of drone audition.

Monaural speech enhancement, or single-channel noise suppression, has been extensively studied for decades.
Traditional approaches include spectral subtraction~\cite{cohen2003noise} and wiener filtering~\cite{haykin2002adaptive}, while recent advances use deep neural networks (DNNs). 
Particularly, masking-based DNN methods showcased promising results in recent Deep Noise Suppression Challenges (DNS)~\cite{dubey2022icassp}. 
These methods often use Convolutional Recurrent Networks (CRNs) to extract features from temporal-spectral patterns, and then predict a ratio mask for noisy speech on the time-frequency spectrum~\cite{tan2018convolutional,hu2020dccrn,zhao2022frcrn}. 
However, these models are typically optimized for scenarios with relatively high input SNRs 
(e.g., $>-5$ dB). 
Applying these pre-trained models directly to drone noise often results in sub-optimal enhancement performance.

Research specifically addresses drone noise suppression is still in its early stages~\cite{wang2020deep}, 
and public drone noise datasets are relatively small~\cite{strauss2018dregon,wang2018acoustic,ruiz2018aira,wang2019audio}, containing only a few types of drones and lacking the diversity needed for training.
Mukhutdinov \textit{et al.}~\cite{mukhutdinov2023deep} comprehensively evaluated the performance of DNNs in monaural speech enhancement on drones.
They rely on limited drone noise datasets, leading to the overfitting to specific drone types.

To address challenges in drone audition, we propose a frequency domain bottleneck adapter for transfer learning, specifically designed to capture the harmonic nature of drone noise.
This adapter-based tuning method selectively trains adapter parameters while retaining the pre-trained model's parameters~\cite{rebuffi2017learning, houlsby2019parameter}, which prevents overfitting with the small-scale training data. 
We apply this method by fine-tuning a pre-trained FRCRN on drone noise datasets.
The proposed method efficiently adapts to the distinct acoustic characteristics of various drone types, thereby enhancing speech quality and intelligibility in drone recordings. 
This advancement paves the way for broader applications of drone audition.

\section{PROBLEM FORMULATION}

\begin{figure}[t]
\centering
\subfloat[]{\includegraphics[width=0.49\linewidth]{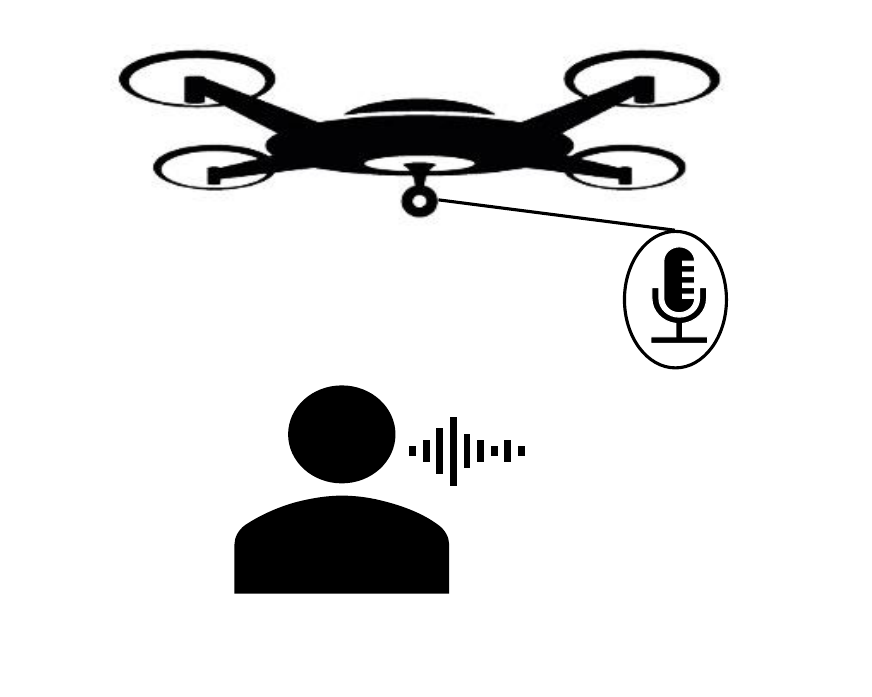}%
\label{figure1a}}
\subfloat[]{\includegraphics[width=0.49\linewidth]{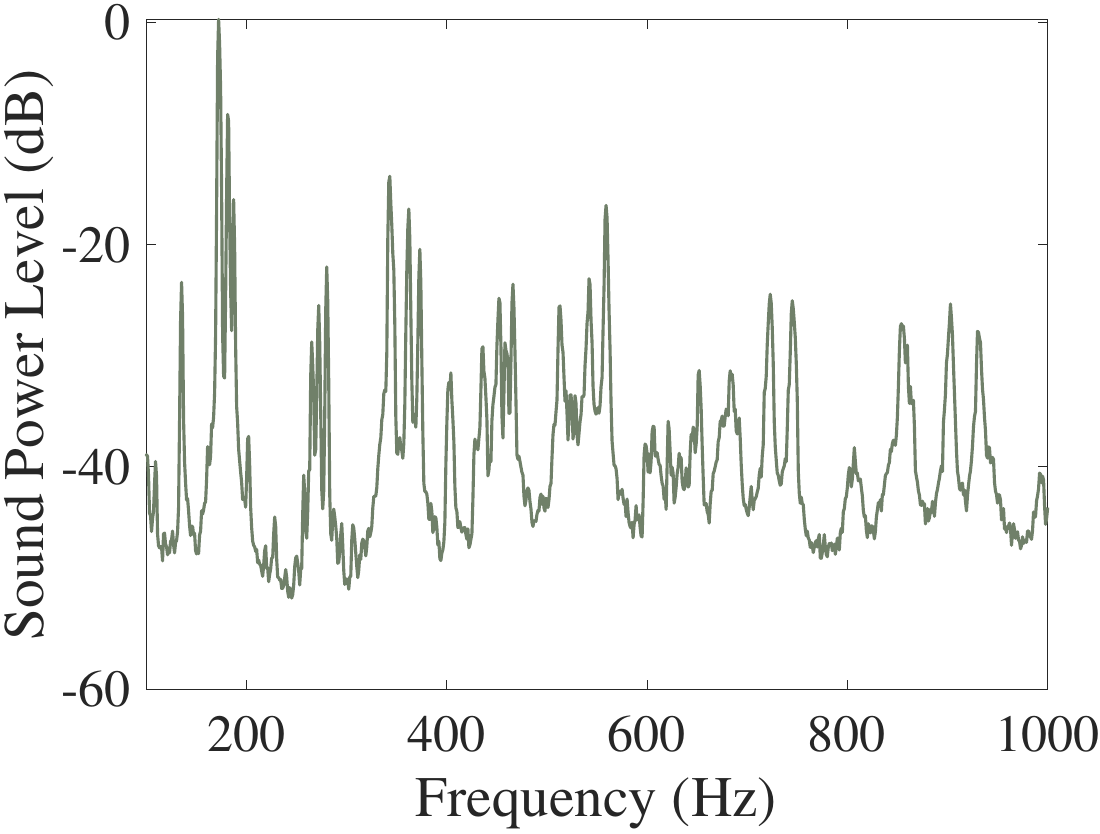}%
\label{figure1b}}

\hfill
\subfloat[]{\includegraphics[width=0.49\linewidth]{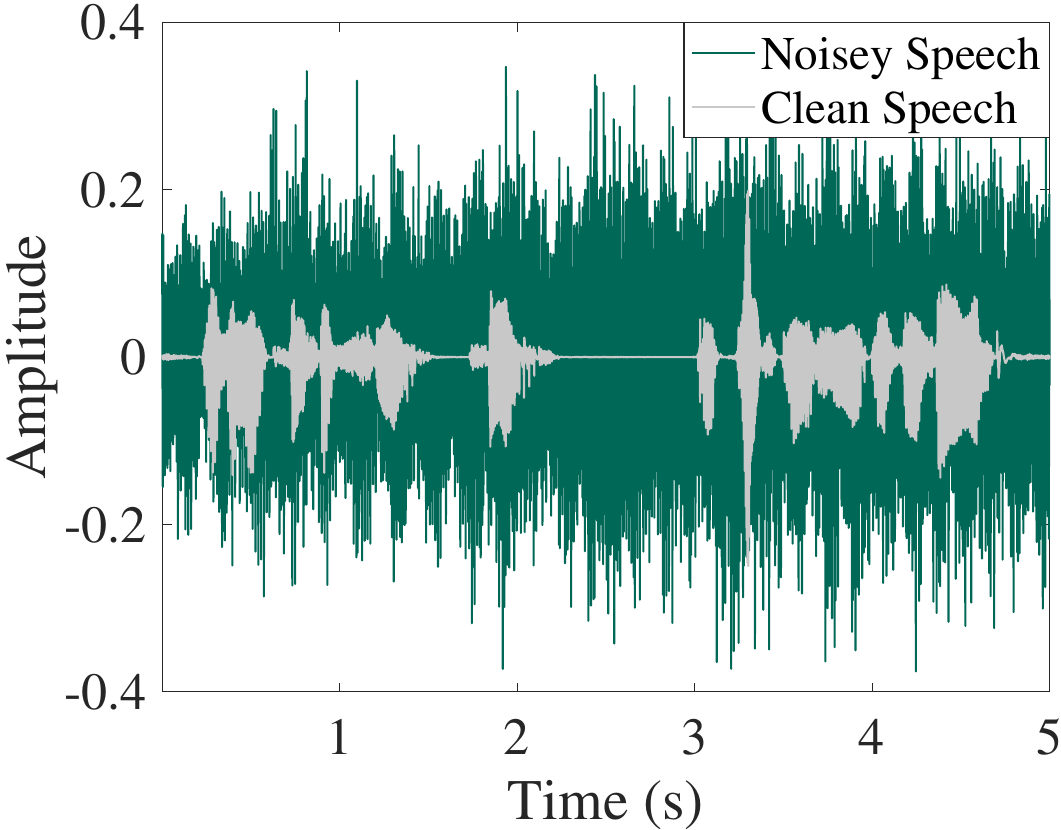}%
\label{figure1c}}
\subfloat[]{\includegraphics[width=0.49\linewidth]{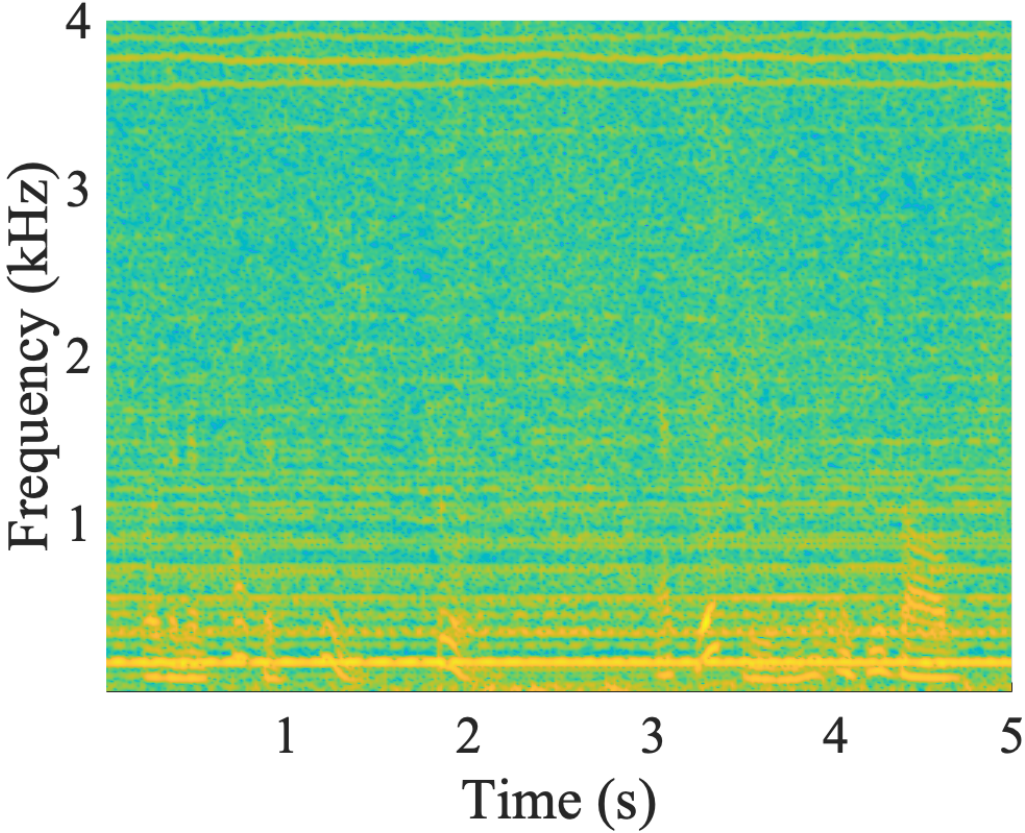}%
\label{figure1d}}

\caption{Problem setup: (a) illustration of the monaural speech recording scenario over a flying drone (b) drone ego noise in the frequency domain (c) noisy speech in the time domain (d) noisy speech in the time-frequency domain.}
\label{Problem formulation}
\end{figure}

Consider a single microphone mounted on a flying drone to capture human speeches as shown in Fig.~\ref{figure1a}. 
The microphone recording can be represented as 
\begin{equation}
y(t)=s(t) * h(t)+v(t),
\label{eq:noisy_t}
\end{equation}
where $s(t)$ denotes the clean speech with $t$ being the time index, $h(t)$ is the impulse response between the human and the microphone, $v(t)$ denotes the noise, and $*$ denotes the convolution operation. $v(t)$ is mainly contributed by the drone propeller rotations and vibrations of the structure and is dominated by multiple narrow-band noises \cite{sinibaldi2013experimental,bi2021spherical}, as demonstrated in Fig.~\ref{figure1b}. As the drone-mounted microphone is close to the noise sources (i.e. motors and propellers), the SNR of $y(t)$ is very low, resulting in the extreme difficulty in uncovering $s(t)$ from $y(t)$. This is demonstrated in Fig.~\ref{figure1c}.

To leverage information from both the time and frequency domains, we apply the Short-Time Fourier Transform (STFT) to Eq.(\ref{eq:noisy_t}), yielding the Time-Frequency (T-F) domain representation:
\begin{equation}
Y(k, l)=S(k, l) H(k, l)+V(k, l),
\label{eq:noisy_t_tf}
\end{equation}
where $Y(k, l), S(k, l)$, and $V(k, l)$ are the complex spectra of $y(t), s(t)$, and $v(t)$, respectively, with $k$ denotes the frequency bin index and $l$ denotes the frame index. 
Fig.~\ref{figure1d} illustrates a typical $Y(k, l)$, highlighting the time-frequency sparsity of harmonic noise and energy concentrated at isolated frequency bins. 

The noisy speech can be enhanced by Complex masking~\cite{williamson2015complex}. In the ideal case, the enhanced speech is obtained by
\begin{equation}
\hat{S}(k, l)=c \operatorname{IRM}(k, l) \odot Y(k, l),
\label{eq:noisy_t_tf}
\end{equation}
where $c \operatorname{IRM}(k, l)$ is the complex Ideal Ratio Mask (cIRM), and 
$\odot$ is element-wise complex multiplication. The cIRM is formulated by
\begin{equation}
c \operatorname{IRM}(k, l)=M_\mathrm{r}(k, l)+i M_\mathrm{i}(k, l),
\end{equation}
where $M_\mathrm{r}(k, l)$ and $M_\mathrm{i}(k, l)$ are, respectively, given by
\begin{equation}
\begin{gathered}
M_\mathrm{r}(k, l)=\frac{Y_\mathrm{r}(k, l) S_\mathrm{r}(k, l)+Y_\mathrm{i}(k, l) S_\mathrm{i}(k, l)}{Y_\mathrm{r}^2(k, l)+Y_\mathrm{i}^2(k, l)}, \\
M_\mathrm{i}(k, l)=\frac{Y_\mathrm{r}(k, l) S_\mathrm{i}(k, l)-Y_\mathrm{i}(k, l) S_\mathrm{r}(k, l)}{Y_\mathrm{r}^2(k, l)+Y_\mathrm{i}^2(k, l)} ,
\end{gathered}
\end{equation} 
where $_r$ and $_i$ denote the real and imaginary components of the spectra, respectively. Since $S(k, l)$ is unknown, directly deriving $c \operatorname{IRM}(k, l)$ is not feasible. 
Then, the research problem reduces to estimate a complex mask $\hat{M}(k, l)$ that approximates $c \operatorname{IRM}(k, l)$ using the single channel recording and the pre-knowledge about drone noise.

\begin{figure*}[t]
\begin{center}
\includegraphics[width=16cm]{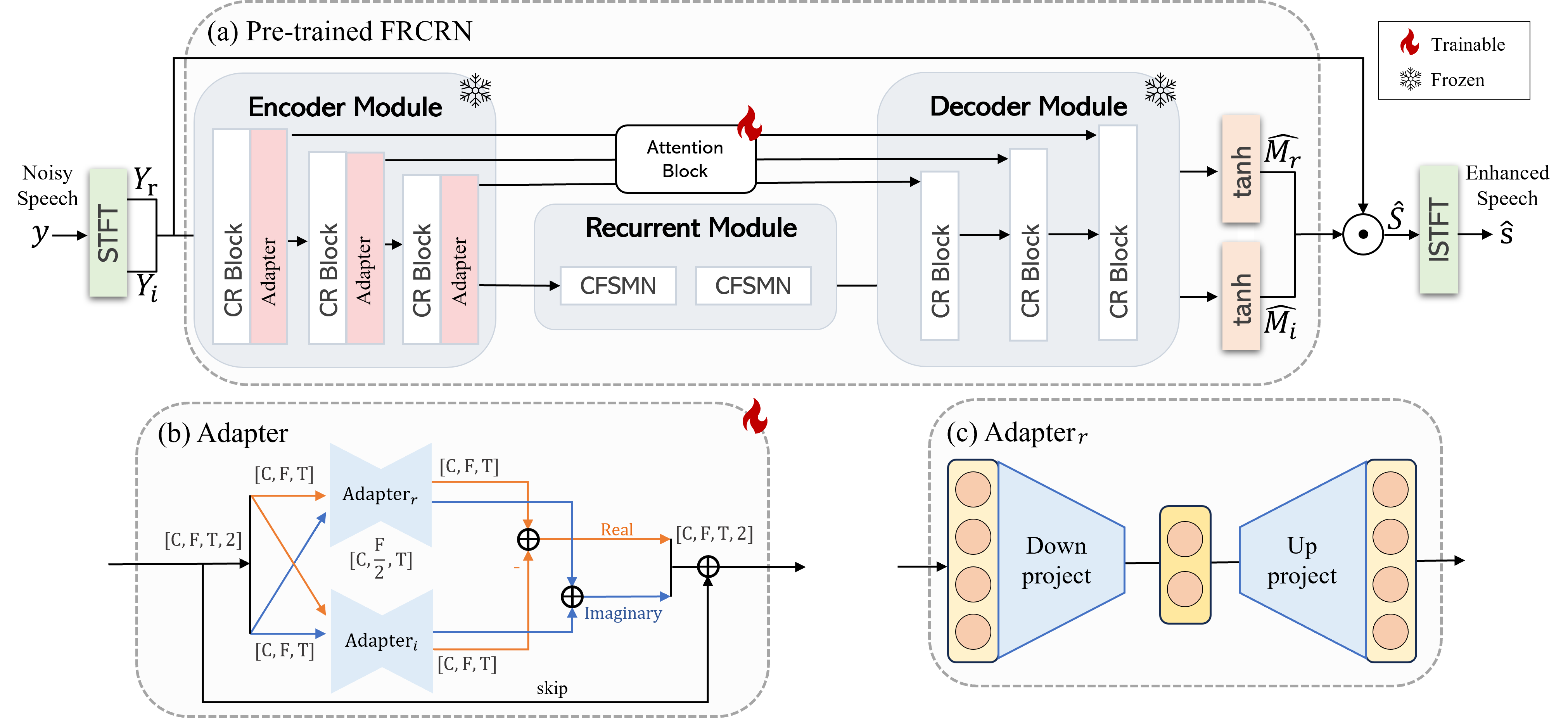}
\end{center}
\caption{The overview of Adapter pipeline. (a) Pre-trained FRCRN with adapter embedded; (b) Adapter; (c) $\mathrm{Adapter}_r$.}
\label{architecture}
\end{figure*}

\section{methodology}

\subsection{Transfer learning with FRCRN}

To enhance monaural speech for drones, we develop a transfer learning-based method 
to estimate the $\hat{M}(k, l)$ using the FRCRN.
FRCRN achieves state-of-the-art (SOTA) results in the recent DNS challenge~\cite{dubey2022icassp}, 
which includes a wide variety of noise types. 
Fig.~2a shows the FRCRN architecture with convolutional recurrent (CR) blocks, 
each comprised of a convolutional layer and a Feedforward Sequential Memory Network 
(FSMN) layer.
This design enables the model to capture local temporal-spectral structures and 
long-range frequency dependencies, making it well-suited for exploiting the 
long-range frequency correlations found in drone noise~\cite{sinibaldi2013experimental}.

\subsection{Adapter tuning on FRCRN}

Transfer learning includes fine-tuning and adapter tuning, 
both of which involve copying the parameters from a pre-trained FRCRN
and tuning them on the target data. 
The fine-tuning trains on a subset of pre-trained model parameters, while the adapter-based 
tuning method only trains on the adapter parameters but keeps the pre-trained model's 
parameters fixed, making adapter tuning more parameter-efficient.

We propose a frequency domain bottleneck adapter 
to learn drone noise characteristics.
Figure~\ref{architecture}a shows the adapter embedded in the encoder module, 
positioned after each CR block.
When tuning the drone dataset, the pre-trained model's parameters are frozen,  
and the parameters of the adapter and the attention block are fine-tuned to learn. 
The attention block acts as the skip pathway to facilitate information flow, hence it is kept unfrozen.

Figure~2b illustrates the structure of the adapter.
The adapter is used to process features in the frequency domain.
The input and output dimensions of the adapter remain consistent.
The adapter has a skip connection, 
and its parameters are initialized to zero, 
configuring the adapter as an approximate identity function.

The detailed operations of the adapter are as follows. 
The adapter consists of two cells for the real and imaginary parts of the input, 
and the outputs are combined according to the properties of complex numbers:
\begin{equation}
\begin{gathered}
\mathbf{A}_\mathrm{r}^{\text{out}}  =\mathrm{Adapter}_\mathrm{r}\left(\mathbf{A}_\mathrm{r}^{\text{in}}\right)-\operatorname{Adapter}_\mathrm{i}\left(\mathbf{A}_\mathrm{i}^{\text{in}}\right), 
\\
\mathbf{A}_\mathrm{i}^{\text{out}} =\mathrm{Adapter}_\mathrm{r}\left(\mathbf{A}_\mathrm{i}^{\text{in}}\right)+\operatorname{Adapter}_\mathrm{i}\left(\mathbf{A}_\mathrm{r}^{\text{in}}\right).
\end{gathered}
\end{equation}

Figure~2c illustrates the real cell operation. For a real part $U_r\in \mathbb{R}^{C \times F \times T}$ of a feature map $U\in \mathbb{R}^{C \times F \times T \times2}$, $C$, $F$, $T$ denote channel, frame, and frequency dimensions, respectively, 
applying the real cell of the adapter $\mathrm{Adapter}_r$ results in the output $U_r'\in \mathbb{R}^{C \times F \times T}$:
\begin{equation}
\begin{aligned}
&\mathbf{h} = \delta(\mathbf{W}_{1} U_r + \mathbf{b}_1),  \\
&U_r' = \mathbf{W}_{2} \mathbf{h} + \mathbf{b}_2.
\end{aligned}
\end{equation}
Here, $\delta$ represents the ReLU activation function. 
$\mathbf{W}_{1}$ performs a frequency domain downward projection, reducing $F$ to $F/2$, 
and $\mathbf{W}_{2}$ is a frequency domain upward projection, expanding $F/2$ back to $F$. 
The terms $\mathbf{b}_1$ and $\mathbf{b}_2$ are biases.
Complex cell shares the same structural as the real cell.

\subsection{Loss function}
The original loss function for FRCRN combines scale-invariant SNR (SI-SNR) and the mean squared error (MSE) losses of cIRM estimates.
In our approach, we use only SI-SNR as the loss function to avoid the need to balance two losses.
The SI-SNR loss $\mathcal{L}(s, \hat{s})$ is defined as~\cite{le2019sdr}.



\section{EXPERIMENTS}


\subsection{Dataset}
We use clean speech and drone ego-noise to generate clean-noisy pairs for training, validation, and testing.
Clean speech data is sourced from DNS-2022~\cite{dubey2022icassp} and LibriSpeech~\cite{panayotov2015librispeech}.
Table 1 details the drone ego-noise datasets, which include AS~\cite{wang2018acoustic}, AVQ~\cite{wang2019audio}, DREGON~\cite{strauss2018dregon} and samples using DJI Phantom 2.
The IDs for AS and AVQ are based on data entries from a public repository\footnote{https://zenodo.org/records/4553667}, while the IDs for DREGON and DJI describe flight states. 
The noise types are categorized based on flight conditions, 
specifically constant denotes noise levels are relatively stable,
dynamic denotes noise levels varying.
For multichannel recordings, only single-channel data is used. 
Some audio samples have been trimmed to remove takeoff sequences, 
and all audio samples are resampled at 16 kHz.

\begin{table}[t]
\centering
\caption{Drone ego-noise dataset}
\begin{tabular}{llll}
\hline
\textbf{Dataset} & \textbf{ID} & \textbf{Noise type} & \textbf{Length [s]} \\ \hline

AS~\cite{wang2018acoustic}& n121 & constant  & 130 \\
 & n122 & dynamic & 140 \\ \hline

 AVQ~\cite{wang2019audio} & n116 & constant  & 120 \\
 & n117 & constant  & 120 \\
 & n118 & constant  & 40 \\
 & n119 & constant  & 210 \\
 & n120 & dynamic & 214 \\ \hline
 
 DREGON~\cite{strauss2018dregon} & Free Flight & dynamic  & 72 \\
 & Hovering & constant & 25 \\ 
  & Up$\&$Down & dynamic & 28 \\ 
   & Rectangle  & dynamic & 25 \\ 
   & Spinning  & constant & 23 \\ \hline
DJI   & Free Flight & dynamic  & 60 \\
 & Hovering & constant & 60 \\ \hline
\end{tabular}
\end{table}

The training set is generated using clean speech from DNS-2022 and drone ego-noise from AVQ (excluding n116), DREGON, and DJI. 
The validation set is generated using clean speech from DNS-2022 and noise from AVQ's n116. 
The test set is created with clean speech from LibriSpeech and drone ego-noise from AS.
Clean speech segments and noise segments are randomly cropped to the same length and then mixed at an SNR varying from -25 to  -5 dB. 
In total, we generate 5 hours of training data, 1 hour of validation data, and 1 hour of testing data. 
The average SNR is -15 dB.
Considering that the duration of existing drone ego-noise datasets are short, 
we did not generate longer-duration data.

\subsection{Evaluation}

We compare the proposed method (\textbf{Adapter tuning}) with 3 methods: 
(i) a pre-trained FRCRN without tuning (\textbf{w/o tuning}); 
(ii) an untrained FRCRN, trained on the training data (\textbf{w/o pre-trained}); 
(iii) a fine-tuning method (\textbf{Fine-tuning}) where only the FSMN in the encoder module and attention block are trainable. 
To conduct a comprehensive evaluation, multiple evaluation metrics are used, including the Perceptual Evaluation of Speech Quality (PESQ)~\cite{rix2001perceptual}, Extended Short-Time Objective Intelligibility Measure (ESTOI)~\cite{jensen2016algorithm}, and SI-SNR~\cite{le2019sdr}. 
Additionally, we compare the efficiency of different methods in terms of the number of trainable parameters.

\begin{figure}[!t]
\centering
\subfloat[]{\includegraphics[width=0.49\linewidth]{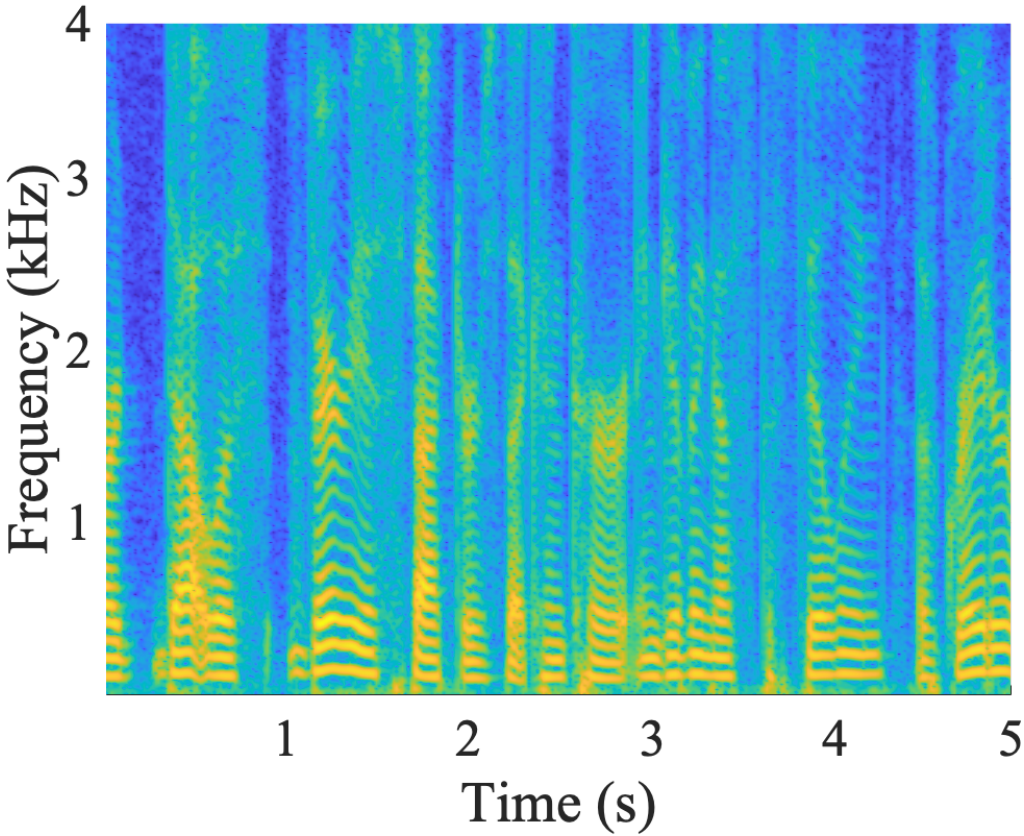}%
\label{figure3a}}
\hfill
\subfloat[]{\includegraphics[width=0.49\linewidth]{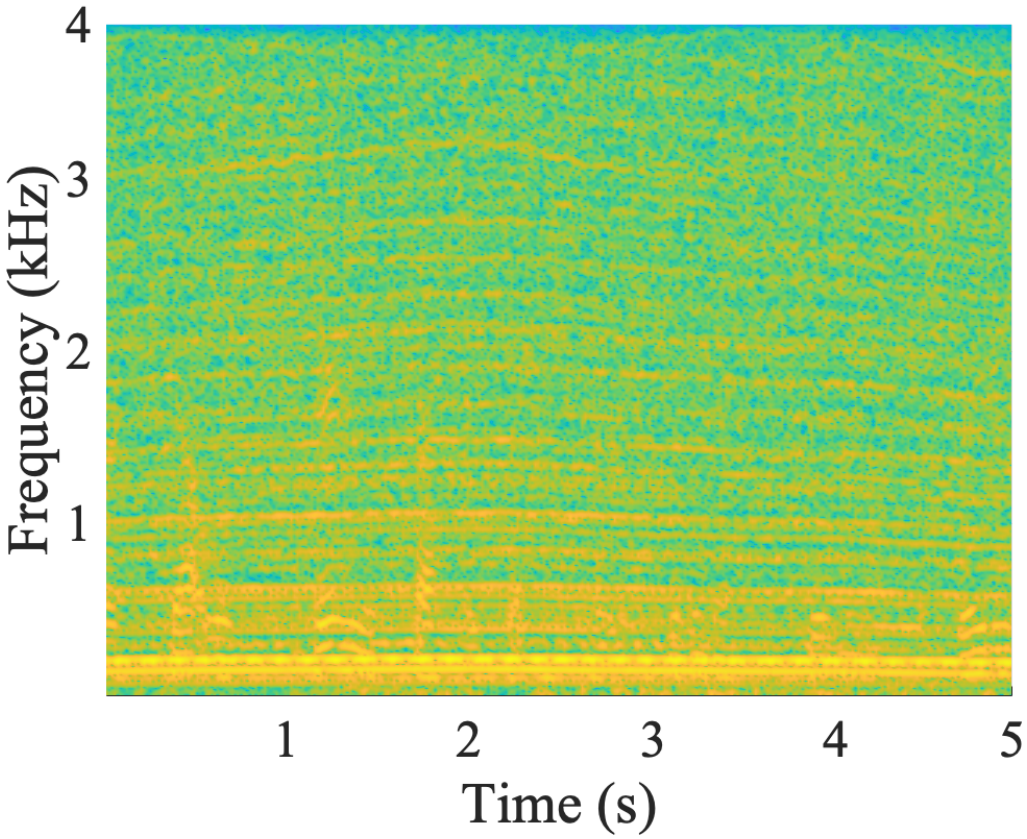}%
\label{figure3b}}

\hfill
\subfloat[]{\includegraphics[width=0.49\linewidth]{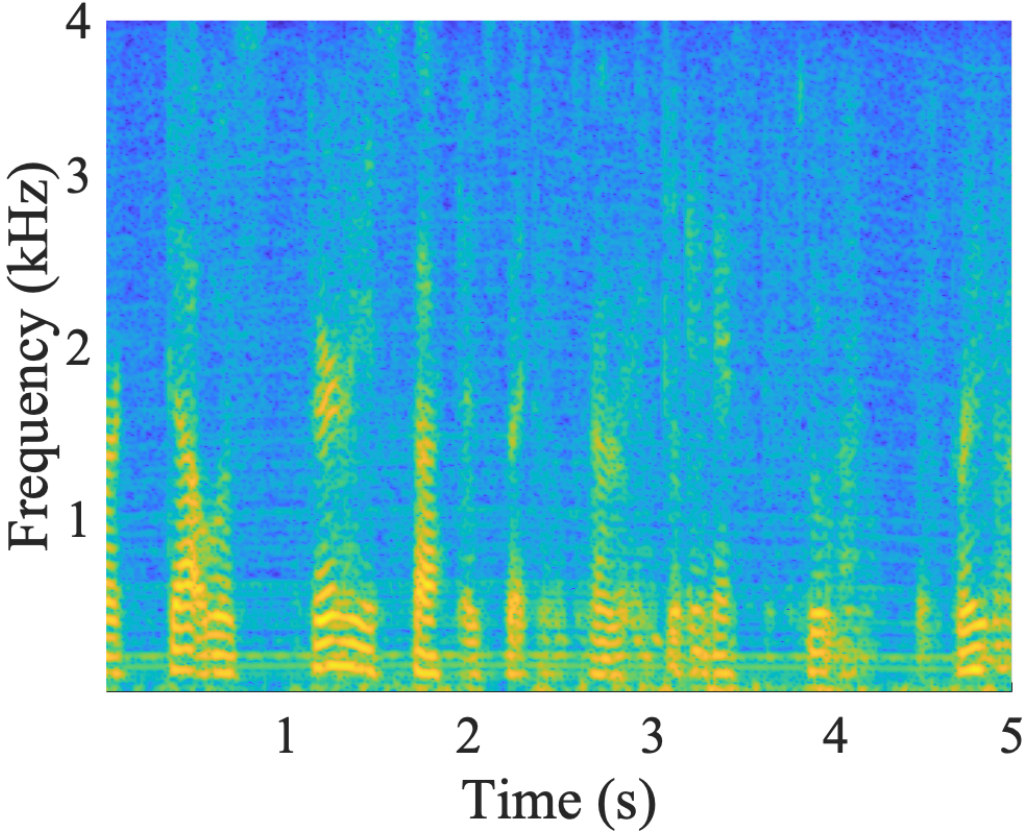}%
\label{figure3c}}
\hfill
\subfloat[]{\includegraphics[width=0.49\linewidth]{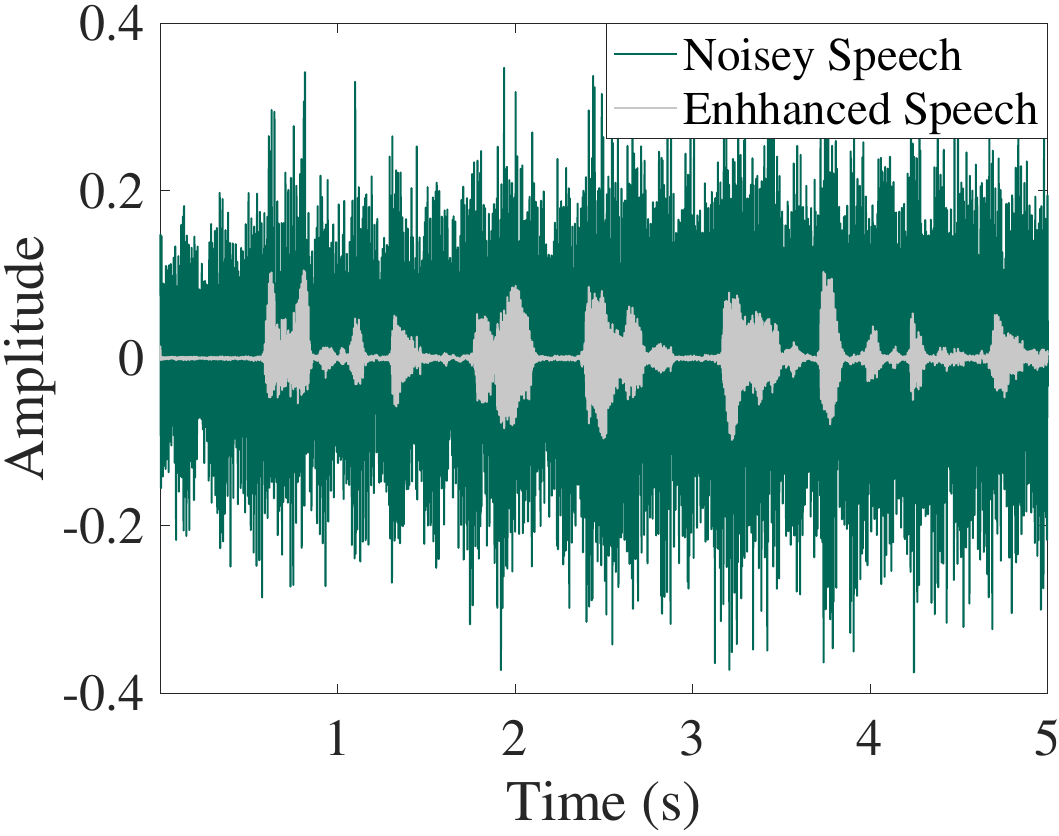}%
\label{figure3d}}
\caption{Results comparison: (a) clean speech (b) noisy speech (c) adapter tuning enhanced speech (d) noisy speech and adapter tuning enhanced speech in time-domain.}
\label{Result}
\end{figure}

\begin{table}[t]
\centering
\caption{Evaluation results}
\begin{tabular}{lccccc}
\hline 
\multirow{2}{*}{} 
&\multirow{2}{*}{\makecell{Trainable\\Params(m)}}
& \multicolumn{3}{c}{Evaluation Metrics} \\
\cline{3-5} 
& & PESQ & ESTOI & SI-SNR \\
\hline 
Noisy speech & - & 1.06 & 0.09 & -14.65 \\
\hline 
w/o tuning & - & 1.04 & 0.32 & -8.49 \\
w/o pre-trained & 14 & 1.25 & 0.36 & 2.56 \\
\hline 
Fine-tuning  & 1.2 & 1.12 & 0.32 & 2.64 \\
Adapter tuning & $\mathbf{0.3}$ & $\mathbf{1.31}$ & $\mathbf{0.47}$ & $\mathbf{2.87}$ \\
\hline 
\end{tabular} 
\end{table}

The results are presented in Table 2. 
Overall, \textbf{Adapter tuning} outperforms the competing methods for all metrics.
The \textbf{w/o tuning} shows limited effectiveness, as it is not designed for drone scenarios.
\textbf{w/o pre-trained} demonstrates improved performance, especially in PESQ and SI-SNR. 
This indicates the significant difference between drone noise and the noise in the normal training dataset.
Although the FSMN layer in the encoder module processes frequency-domain information as the proposed adapter, 
\textbf{Fine-tuning} does not yield optimal results.
This may be because \textbf{Fine-tuning} leads to the forgetting of previously learned information, whereas \textbf{Adapter tuning} preserves all pre-trained parameters. 
The numbers of trainable parameters indicate the efficiency of the proposed method which surpasses \textbf{Fine-tuning} with much less trainable parameters.


Figure 3 illustrates the time-frequency spectra of a segment of clean speech, noisy speech, adapter enhanced speech, and the time-domain plot of the speech before and after enhancement. 
As shown in Fig.~\ref{figure3b}, (i) in the noisy speech, the speech is obscured in the drone noise, and sound power is mainly concentrated in the harmonic frequencies; 
(ii) comparing Fig.~\ref{figure3c} with Fig.~\ref{figure3a}, the enhanced speech has a similar sound power distribution to the clean speech, indicating the effectiveness of the proposed method; 
(iii) as shown in Fig.~\ref{figure3d}, the enhanced speech is much cleaner with an 18 dB SNR 
improvement. 









\section{conclusions}
Due to the cost and weight constraint, monaural speech enhancement for drones 
is preferred over multichannel speech enhancement. However, the absence of spatial 
information and the low SNR makes monaural speech enhancement challenging. 
By exploiting the harmonic nature of the drone ego noise, this paper developed a
frequency domain bottleneck adapter for transfer learning.  
The method takes advantage of transfer learning to re-uses knowledge learned from large data, 
thereby achieving effective training on small data. 
The proposed method enhances speech quality and intelligibility in drone recordings, paving 
the way for broader drone audition applications.

\clearpage
\bibliographystyle{IEEEbib}
\bibliography{refs}

\end{document}